\documentclass[pre,preprint,showpacs]{revtex4}
\usepackage{graphicx,amssymb}

\begin{document}

\title{Soliton Lattice and Single Soliton Solutions of the Associated Lam\'e and
Lam\'e Potentials}

\author{Ioana Bena$^\ast$, Avinash Khare$^\dag$, and Avadh Saxena$^\ddag$}
\address{$^\ast$Department of Theoretical Physics, University of Geneva, CH-1211, 
Geneva 4, Switzerland\\  
$^\dag$Institute of Physics, Bhubaneswar, Orissa 751005, India\\  
$^\ddag$Theoretical Division, Los Alamos National Laboratory, Los Alamos, 
New Mexico 87545, USA 
}

\date{\today}

\begin{abstract}
{We obtain the exact nontopological soliton lattice solutions  of the 
Associated Lam\'e equation in different parameter regimes and compute 
the corresponding energy for each of these solutions. We show that   
in specific limits these solutions give rise to 
nontopological (pulse-like) single solitons, as well as to different types of 
topological (kink-like) single soliton solutions of the Associated Lam\'e equation. 
Following Manton, we also compute, as an illustration, the asymptotic 
interaction energy between these soliton solutions in one particular case. 
Finally, in specific limits, we deduce the soliton 
lattices, as well as the topological single soliton solutions 
of the Lam\'e equation, and also the sine-Gordon soliton solution.} 

\end{abstract}


\maketitle

\section{Introduction}

Over the years, extensive research has been carried out seeking the 
exact soliton solutions of both periodic (e.g., sine-Gordon, double 
sine-Gordon) and nonperiodic (e.g., $\phi^4$, $\phi^6$) field theory 
models.  For example, the exactly solvable sine-Gordon (SG) 
equation \cite{sg} and its quasi-exactly solvable (QES) partner, 
i.e., the double sine-Gordon equation (DSG), have exact single soliton 
\cite{leung,peyrard} as well as soliton lattice \cite{dando} solutions.
  
There have been some advances in the study of the {\em hyperbolic} analogues 
of these problems; the exactly solvable 
hyperbolic analogue of the SG equation is the sine-hyperbolic Gordon 
(ShG) equation \cite{shg}. This potential has only one minimum and 
thus does not support (topological) soliton solutions.  The hyperbolic 
analogue of the DSG equation is the double sine-hyperbolic Gordon 
(DShG) equation, which is a QES double-well potential with exact single soliton 
and soliton lattice solutions \cite{dshg}. 

However, not much is known regarding the {\em elliptic} analogues 
of these problems. 
The elliptic `generalization' of the SG is the Lam\'e equation 
\cite{arscott} but, as far as we are aware of, its single soliton and soliton 
lattice solutions have not been worked out yet.  
One would surmise that the elliptic analogue of the DSG equation is 
the Associated Lam\'e (AL) equation \cite{magnus,khare,ganguly}. 
Nevertheless, we find below that this is not the case. 

Generally speaking, the Schr{\"o}dinger equations with periodic potentials 
belong to a class  known 
as Hill's equations \cite{magnus}, and they lead 
to a band structure of the energy spectrum of the system.
For both the SG and DShG potentials one could calculate the exact 
statistical and thermodynamical properties \cite{dshg,gupta} 
through a knowledge of the spectral band structure and their solitonic 
and phononic solutions, respectively.  
Both AL and Lam\'e  are also periodic potentials with 
a band structure \cite{arscott,khare,ganguly}.  Although our goal in the present 
paper is only to obtain their exact single soliton and soliton lattice solutions, 
this will serve as a background for future statistical mechanics 
studies of the AL and Lam\'e elliptic potentials. In particular, 
we obtain here the soliton lattice solutions of the AL equation in 
different parameter regimes 
corresponding to different shapes of the AL potential. 
The advantage of this approach is
that the corresponding single pulse and topological soliton solutions
of the AL equation can be immediately obtained from the lattice 
solution in suitable limits. Furthermore, the soliton lattices and 
topological single soliton solutions of the Lam\'e equation are easily 
obtained by taking a different appropriate limit. 
Besides, the 
asymptotic interaction between the solitons can also be readily 
obtained by using Manton's formalism \cite{manton}. 

Our results can be summarized as follows: We show that there are six 
different soliton lattice solutions of the AL equation for different 
values of the parameters. From these six solutions we obtain the  
corresponding AL single soliton solutions. It turns out that while in 
five cases these are topological (kink-like) single solitons, in one 
case we have a nontopological (pulse-like) single soliton solution. 
From the AL soliton lattice solutions, by taking appropriate limits, 
we also obtain the corresponding Lam\'e soliton lattices and single 
soliton solutions.

There are many physical contexts in which the Lam\'e equation arises, 
such as bond-order and charge density wave systems \cite{horovitz}, 
nonlinear elasticity \cite{zhurav}, and other contexts, e.g., phase 
slips in superconductors and magnetoelastic interaction on curved 
surfaces.  The AL equation could arise in the above systems in the 
presence of external  -- electric, magnetic or stress -- fields.  

The plan of the paper is as follows. In Sec. II we  discuss
some salient features of the AL potential.  In Sec. III we  obtain 
the nontopological soliton lattice solutions of the AL equation 
in different regimes in the parameter space 
and also compute the corresponding energies. 
In Sec. IV we obtain the
topological and pulse single soliton solutions of the AL equation by taking
the appropriate limits of the various nontopological 
solutions obtained in the previous
section. It is worth emphasizing here that in one special case, we obtain
two different kinds of topological solutions of the AL equation. As an
illustration, in one particular case, we also 
estimate the asymptotic interaction between these single solitons.    
(In all the other cases one can follow exactly the same procedure, 
and thus we will not consider these here.) 
As a crosscheck on our results, we recover in appropriate 
limit, the sine-Gordon soliton solution.
In Sec. V we show that by taking an appropriate limit, 
we can also obtain the kink 
lattice and the topological single soliton solutions of the Lam\'e equation.
Finally, in Sec. VI we summarize the results obtained in this paper
and indicate some open problems.

\section{The Associated Lam\'e potential}

Consider the following family of periodic potentials labeled by a pair 
of real parameters $(p,\,q)$:
\begin{equation}
V_{AL}(\phi,\,k)\,=\,p\, k^2\,
\mbox{sn}^2(\phi,\,k)\,+\,q\,k^2\,\displaystyle\frac{\mbox{cn}^2(\phi,\,k)}
{\mbox{dn}^2(\phi,\,k)}\,+\,C\,=
\,p\, k^2\,
\mbox{sn}^2(\phi,\,k)\,+\,q\,k^2\,\displaystyle{\mbox{sn}^2(\phi+K(k),\,k)}\,+\,C~,
\label{assocLame}
\end{equation}
that are called {\em Associated Lam\'e potentials} (since the corresponding
Schr\"odinger equation is called the Associated Lam\'e equation) 
\cite{magnus,khare,ganguly}.
Here $\mbox{sn}(\phi,\,k)$ and $\mbox{cn}(\phi,\,k)$ are respectively the 
sine and cosine amplitude Jacobi elliptic functions of real modulus 
$k$ ($0\,\leqslant\,k\,\leqslant 1$) and  period $4\,K(k)$; 
$\mbox{dn}(\phi,\,k)$ is the $\delta$-amplitude Jacobi elliptic function of
modulus $k$ and period $2\,K(k)$; and $K(k)$ denotes the complete elliptic 
integral of the first kind, see \cite{gradshteyn,byrd}.
We will choose  the constant $C$ in the potential so that
the absolute  minimum of the potential--with respect to $\phi$--equals zero,  
$V_{min}\,=\,0$.  The potential is periodic with period $2K(k)$, except in 
the limit $p=q$ when the period reduces to K(k), 
as it is clear from Eq.~(\ref{assocLame}). 

The case $q\,=\,0$ corresponds to the standard Lam\'e potential 
$V_L(\phi,\,k)$.  There are two more cases in which the potential 
(\ref{assocLame}) reduces to the Lam\'e potential. Namely, when $p\,=\,0$  
and also when $p\,=\,q$, as shown below. Therefore in all the calculations  
below we will admit that $p\,\neq\,0$ and $p\,\neq\,q$; the 
results for the standard Lam\'e potential will be recovered in the limit 
$q\,\rightarrow 0$, and we refer to these results whenever the potential
reduces to the standard Lam\'e one. From a physical point of view, if one thinks
of a Lam\'e potential $(p,0)$ as due to a one-dimensional array of atoms with
spacing $2K(k)$ and `strength' $p$, then the Associated Lam\'e potential $(p,q)$
results from two alternating types of atoms spaced by $K(k)$ with `strengths'
$p$ and $q$, respectively. If the two types of atoms are identical (which makes
$p~=~q$), one expects a potential of period $K(k)$. 

In addition, when $k\,\rightarrow \,0$, and $|p|,\,|q|\,\rightarrow \infty$, 
so that 
$|p|\,k^2\,\rightarrow\, P$ = finite and $|q|\,k^2\,\rightarrow\, Q$ = finite, 
the potential (\ref{assocLame}) reduces to the sine-Gordon potential,
\begin{equation}
V_{sG}\,=\,(P\;\mbox{sign}\;p\,-\,Q\;\mbox{sign}\;{q})\,\sin^2\phi\,,
\label{sG}
\end{equation}
and  the results we present below reduce  to the well-known 
ones for the sine-Gordon potential.

It is also worth noting that under the transformation $\phi \rightarrow \phi+K(k)$
the AL potential~(\ref{assocLame}) $V_{AL}(p,q)$ goes over into $V_{AL}(q,p)$ and hence
for $p > 0, q \ge 0$, as well as for $p <0, q \le 0$, without loss of 
generality, we shall always consider the case of $p^2 > q^2$. Further, 
instead of considering both the possibilities of $p > 0, q \le 0$ and
$p < 0, q \ge 0$, it suffices to consider just the case of $p > 0, q \le 0$,
but now $p^2$ can be bigger as well as smaller than $q^2$. 

Finally, consider the case of $p=q$, when, by using the Landen transform
\cite{gradshteyn}
and choosing the constant $C=-pk^2$, the AL potential (\ref{assocLame}) can 
be written as
\begin{equation}
V_{AL} (\phi,k)=(1-k')^2V_{L} \left[(1+k')\phi,\frac{1-k'}{1+k'}\right]~,
\end{equation}
where $k'=\sqrt{1-k^2}$ is the complementary elliptic modulus.   
For the rescaled field $\tilde{\phi}=(1+k')\phi$ and in the rescaled 
space coordinate $\tilde{x}=(1+k')x$ the field equations [see next section, 
Eq. (\ref{stateq})] will remain the same as that for the simple Lam\'e 
potential. Note that the $p=q$ case cannot be obtained as a limit of $q 
\rightarrow p$ ($q\neq p$), since--as already mentioned--when $p=q$ the 
periodicity of the potential (\ref{assocLame}) is $K(k)$, 
while for $p\neq q$ it is $2\,K(k)$, see also \cite{khare}.

\section{Soliton Lattice Solutions of the AL Potential}

For the scalar field $\phi\,=\,\phi(x,\,t)$ the dynamics is described by the
second-order hyperbolic differential equation 
\begin{equation}
\displaystyle\frac{\partial^2 \phi}{\partial t^2}\,-\,
\displaystyle\frac{\partial^2\phi}{\partial x^2}\,=\,-\,
\displaystyle\frac{\partial V_{AL}}{\partial \phi}\,.
\label{eveq}
\end{equation}
In the stationary case $\phi\,=\,\phi(x)$, this reduces simply to
\begin{equation}
\displaystyle\frac{d^2\phi}{d x^2}\,=\,
\displaystyle\frac{\partial V_{AL}}{\partial \phi}\,,
\label{stateq}
\end{equation} that can be easily integrated, at least formally, by quadratures.
The time-dependent solutions are immediately obtained from here by Lorentz
boosting.  The physically meaningful solutions 
(bounded at $x\,\rightarrow\,\pm\,\infty$) are given by
\begin{equation}
\pm\,\sqrt{2}\,(x\,-\,x_0)\,=\,\displaystyle\int_{\displaystyle\phi(x_0)}^
{\displaystyle\phi(x)}
\displaystyle\frac{d\tilde{\phi}}{\sqrt{V_{AL}(\tilde{\phi},\,k)\,-\,A^2}}\,,
\label{quadrature}
\end{equation}
where $x_0$  and $A^2\equiv |p| \,k^2\,a^2$ are the two (suitably chosen) integration
constants. Of course, $V_{min}\,=\,0 \leqslant\, A^2 \,<\, V_{max}$, where 
$V_{min/max}$ is the absolute 
minimum/maximum of the potential energy with respect to $\phi$. In fact, in view
of the condition $V_{AL}(\tilde{\phi},\,k)\,-\,A^2 \geqslant 0$,  the actual value of
$A^2$ determines the appropriate integration domain in Eq. (\ref{quadrature}). 

Depending on the values of the parameters $p$ and $q$, the potential 
$V_{AL}(\phi, k)$ has different behaviors in one period
$0\,\leqslant\,\phi\,<\,2\,K(k)$, and hence the nature of the solutions in Eq.
(\ref{quadrature}) is also different.
There are three cases to be considered separately, namely (I) when  $p\,>\,0$
and $q\,\geqslant\,0$,  (II) when $p\,<\,0$ and $q\,\leqslant\,0$, and, finally, 
(III) when $p\,>\,0$ and $q\,\leqslant\,0$. Note that in what follows  
we shall use the shorthand notation $\Gamma\,\equiv\,\sqrt{|{q}|/|{p}|}$.

\subsection{Case I : $p\,>\,0$ and $q\,\geqslant\,0$}
\label{caseI}

As explained above, in this case it is 
sufficient to  consider $0\,\leqslant \,\Gamma\,<\,1$.  Note that the case
$\Gamma\,=\,0$ corresponds to the standard Lam\'e potential. 
Depending on the value of $\Gamma$, the potential $V_{AL}(\phi,\,k)$ can have
different behaviors in one period $0\,\leqslant \,\phi\,<\,2\,K(k)$.

\subsubsection*{Case I.1 : {$0\,\leqslant \,\Gamma\,\leqslant \,k'$}}
\label{caseI.1} 

Here $0\,\leqslant\,k'\,=\,\sqrt{1\,-\,k^2}\,\leqslant\,1$ is the complementary
elliptic modulus. The potential $V_{AL}(\phi)$ has only one minimum, $V_{min}\,=\,0$ (at  
$\phi\,=\,0$), and one maximum, $V_{max}\,=\,p\,k^2\,(1\,-\,\Gamma^2)$ [at
$\phi\,=\,K(k)$],  in one period [note that we have chosen 
$C\,=\,-\,p\,k^2\,\Gamma^2$ in Eq.~(\ref{assocLame}) in order to have $V_{min}=0$]. 
The plot of the potential $V_{AL}$
as a function of $\phi$ is given in Fig.~1(a), where  the 
solid and the dashed horizontal lines correspond 
to two choices of the parameter 
$A^2$. Under the change of variable
$\tilde{z}\,=\,\mbox{sn}^2(\tilde{\phi},\,k)$, 
Eq. (\ref{quadrature}) can be rewritten as
\begin{equation}
2\,\sqrt{2\,p\,k^4}\,x\,=\,\pm\,\displaystyle\int_{\displaystyle z}^
{\displaystyle 1}\,
\displaystyle\frac{d\tilde{z}}{\sqrt{(z_1\,-\,\tilde{z})\,(1\,-\,\tilde{z})\,
(\tilde{z}\,-\,z_2)\,(\tilde{z}\,-\,0)}}\,,
\label{caseI1}
\end{equation}
where  $z=\mbox{sn}^2(\phi,\,k)$, $x_0\,=\,0$ for $\phi_0\,=\,K(k)$ (by 
choice) and  $A^2\, =\,p\,k^2\,a^2$ with 
$0\,\leqslant a^2\,<\,(1\,-\,\Gamma^2)$.  The value of $a^2$ determines 
the limits of the integration interval in the above equation, through the
condition $V(\phi_{1,2})\,-\,p\,k^2\,a^2\,=\,0$; one finds that   
$z_{1,2}\,=\,\mbox{sn}^2(\phi_{1,2})$ are given by:
\begin{equation}
z_{1,\,2}\,=\,\displaystyle\frac{1\,-\,\Gamma^2\,k^{'2}\,+\,a^2\,k^2}{2\,k^2}\,
\left[1\,\pm\,\sqrt{1\,-\,\displaystyle\frac{4\,k^2\,a^2}
{(1\,-\,\Gamma^2\,k^{'2}\,+\,a^2\,k^2)^2}}\;\right]\,,
\label{z12I.1}
\end{equation}
and  $z_1\,>\,1\,>\,z\,\geqslant\,z_2\,>\,0$. Then the integral in (\ref{caseI1})
can be evaluated using the formula 3.147(5) of \cite{gradshteyn}
and one is finally led
to the {\em nontopological} soliton lattice solution  
\begin{eqnarray}
\mbox{sn}^2(\phi,\,k)&=&\displaystyle\frac{[(z_1\,-\,z_2)\,-\,z_1\,(1\,-\,z_2)\,
\mbox{sn}^2(y,\,t)]}{[(z_1\,-\,z_2)\,-\,(1\,-\,z_2)\,\mbox{sn}^2(y,\,t)]}\,.
\label{latticeI.1}
\end{eqnarray}
Here $y\,=\,\sqrt{2\,p\,k^4\,(z_1\,-\,z_2)}\;x$ and the period of the
lattice $2\,L\,=\,2\,K(t)/\sqrt{2\,p\,k^4\,(z_1\,-\,z_2)}$ is controlled by 
the modulus
\begin{equation}
0\,<\,t\,=\,\left[\displaystyle\frac{z_1\,(1\,-\,z_2)}{z_1\,-\,z_2} 
\right]^{1/2}\,
\leqslant\,1.
\label{m}
\end{equation}

Note that $\phi$ oscillates between $K(k)$ and $\mbox{sn}^{-1}(\sqrt{z_2})$
[respectively, $-\,K(k)$ and $-\,\mbox{sn}^{-1}(\sqrt{z_2})$], i.e., it is 
indeed a nontopological solution [it does not connect two adjacent degenerate 
minima of the potential, that are separated by a distance--in the $\phi$ 
space--of $2\,K(k)$].  In Fig. 1(b) we give a plot of the solution (i.e., 
the field $\phi$ as a function of $x$) with solid and dashed curves 
corresponding to the two choices of $A^2$ as in Fig. 1(a), while the dotted 
curve represents the {\em topological} single soliton solution (corresponding 
to $A^2=0$)--see Sec. IV.  We also note that there are no ``cusps"
in the actual profiles of the fields; indeed, in Fig. 
1(b) and in the subsequent figures  we have plotted 
modulo $2K(k)$ the absolute value of the actual field $\phi(x)$.  Specifically, 
for solutions that touch $K(k)$ such as in Fig. 1(b), the actual field is 
$\phi(x)$ in $[0,2L]$, $2K(k)-\phi(x)$ in $[2L,4L]$, $\phi(x)$ in $[4L,6L]$, 
and so on.  Similarly,  for solutions that touch $0$ field value such as in 
Fig. 3(b), the actual field is $\phi(x)$ in $[0,2L]$, $-\phi(x)$ in $[2L,4L]$, 
$\phi(x)$ in $[4L,6L]$, and so on.   

\begin{figure}[htbp]
\begin{center}
\includegraphics[width=0.8\textwidth,angle=-90]{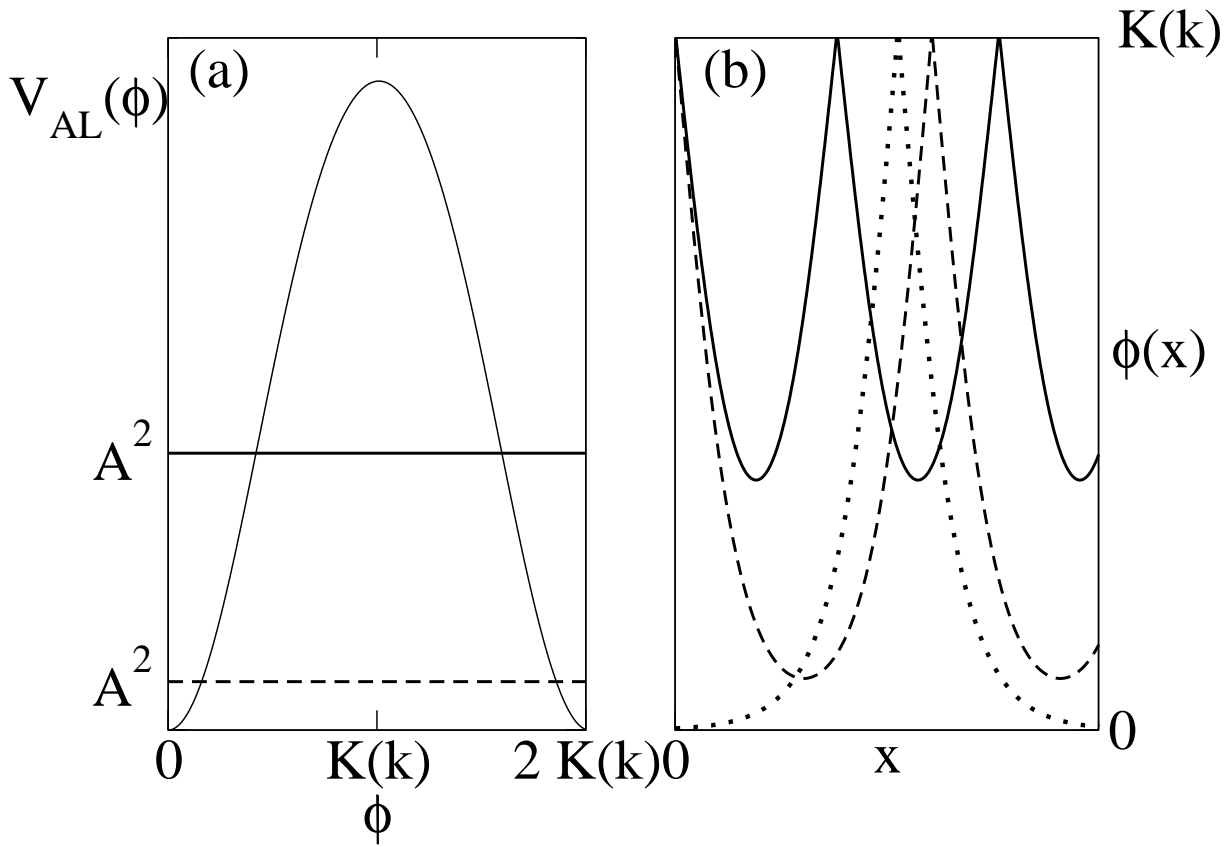}
\end{center}
\caption{Case I.1 : {$0\,\leqslant \,\Gamma\,\leqslant \,k'$}. 
(a) The shape of the potential $V_{AL}(\phi)$.  The solid and the dashed 
horizontal lines correspond to two choices of the integration parameter 
$A^2$. 
(b) The soliton lattice solution [i.e., the field $\phi (x)$] with solid 
and dashed curves corresponding to the two values of $A^2$ in (a). The 
$x$-axis is not labeled because the two lattices have different spatial 
periods.  The dotted curve represents the topological single soliton 
solution corresponding to $A^2=0$ (for the sake of clarity we have 
displaced the origin of the $x$-axis for this curve). }  
\label{fig1}
\end{figure}

One can compute the energy corresponding  to a period $2\,L$ of this 
soliton lattice:
\begin{eqnarray}
E_{SL}&=&\displaystyle\int_{\displaystyle -\,L}^{\displaystyle L}\,\left 
[\frac{1}{2}\;\left(\frac{d^2\,\phi}{dx^2}\right)\,+\,V_{AL}(\phi)\right]\,dx\,=\,
\int_{\displaystyle -\,L}^{\displaystyle L}\,
\left[2\,V_{AL}(\phi)\,-\,A^2\right]\,dx\, \nonumber \\
&=&\,4\,\int_{\displaystyle 0}^
{\displaystyle L} V_{AL}(\phi)\,dx\,- 2\,L\,A^2
=4\,p\,k^2\,\int_{\displaystyle 0}^
{\displaystyle L}\,\mbox{sn}^2[\phi(x),\,k]\,dx\, \nonumber \\
&+&\,4\,p^2\,k^2\,\Gamma^2\,
\int_{\displaystyle 0}^{\displaystyle L}\left\{
\displaystyle\frac{\mbox{cn}^2[\phi(x),\,k]}{\mbox{dn}^2[\phi(x),\,k]}
\,-\,1\right\}\,dx\,-\, 2\,L\,p^2\,k^2\,a^2\,.
\end{eqnarray}
After some algebraic manipulations, using the solution (\ref{latticeI.1})  
one obtains:
\begin{eqnarray}
E_{SL}&=&2\,\sqrt{\displaystyle\frac{2\,p}{z_1\,-\,z_2}}\,
\left\{\left[z_1\,K(t)\,-\,(z_1\,-\,1)\,
\Pi\left(\displaystyle\frac{1\,-\,z_2}{z_1\,-\,z_2},\,t\right)\right]\,-\,
\displaystyle\frac{a^2\,K(t)}{2}\right\}\,-\,
\nonumber\\
&&\nonumber\\
&&-\,\Gamma^2\,\left[\displaystyle\frac{z_1\,k^{'2}}{1\,-\,z_1\,k^2}\,K(t)\,-\,
\displaystyle\frac{z_1\,-\,1}{1\,-\,z_1\,k^2}\,\Pi
\left(\displaystyle\frac{(1\,-\,z_2)\,(1\,-\,z_1\,k^2)}
{(z_1\,-\,z_2)\,k^{'2}},\,t\right)\right]\,,
\label{energyI.1}
\end{eqnarray}
where $\Pi(z_0,t)$ denotes the complete elliptic integral of the third kind
\cite{gradshteyn,byrd}.

\subsubsection*{Case I.2 : {$k'\,<\,\Gamma\,<\,1$}}
\label{caseI.2} 

The potential $V_{AL}(\phi)$  has now
two minima: $V_{min}(\phi\,=\,0)\,=\,0$ and a local minimum $V^l_{min} 
[\phi\,=\,K(k)]\,=\,p\,k^2\,(1\,-\,\Gamma^2)$, and two symmetric maxima 
around $\phi~=~K(k)$, namely
$V_{max}\,=\,p\,k^2\,[(1\,-\,\Gamma\,k')/k]^2$ for $\phi\,=\,
\mbox{sn}^{-1}\left[\sqrt{(1\,-\,\Gamma\,k')/k^2}\right],\,
2\,K(k)\,-\,\mbox{sn}^{-1}\left[\sqrt{(1\,-\,\Gamma\,k')/k^2}\right]$.
The plot of the potential $V_{AL}(\phi)$ as a function of $\phi$ is given in
Fig.~2(a), where the solid and the dashed horizontal lines correspond 
to the two choices of the integration constant $A^2$ as explained below.  Note that 
in this case it makes no sense to consider either the limit of the Lam\'e 
potential (i.e., $\Gamma\,\rightarrow\,0$), or that of the sine-Gordon
potential (i.e., $k\,\rightarrow\,0$,
$p\,k^2\,\rightarrow\,P,\;q\,k^2\,\rightarrow\,Q$).
There are two possible situations, depending on the value of $A^2=p\,k^2\,a^2$.

{\em Case I.2(i)} : {$V_{min}\,\leqslant\,A^2=p\,k^2\,a^2\,<\,V^l_{min}$}, 
i.e., $0\,\leqslant a^2\,<\,1\,-\,\Gamma^2$.\\
One recovers the same soliton
lattice solution as above,  Eq. (\ref{latticeI.1}), 
with the same modulus $t$, Eq. (\ref{m}), 
and the same energy per period of the lattice, 
Eq. (\ref{energyI.1}).

{\em Case I.2(ii)} : {$V^l_{min}\,\leqslant\,A^2=p\,k^2\,a^2\,<\,V_{max}$}, 
i.e., $1\,-\,\Gamma^2\,\leqslant\,a^2\,<\,[(1\,-\,\Gamma\,k')/k]^2$.\\
Then the integral (\ref{quadrature}) becomes 
\begin{equation}
2\,\sqrt{2\,p\,k^4}\,x\,=\,\pm\,\displaystyle\int_{\displaystyle z_2}^
{\displaystyle z}\,
\displaystyle\frac{d\tilde{z}}{\sqrt{
(1\,-\,\tilde{z})\,(z_1\,-\,\tilde{z})\,
(\tilde{z}\,-\,z_2)\,(\tilde{z}\,-\,0)}}\,,
\label{caseI2}
\end{equation}
with $z\,=\,\mbox{sn}^2(\phi,\,k)$ and  $1\,>\,z_1\,\geqslant\,z\,>\,z_2$
[$z_{1,2}$ have the same expressions as above, Eq. (\ref{z12I.1})].
One can evaluate this integral using the formula 3.147(4) of \cite{gradshteyn}  
and obtain the following {\em nontopological} soliton
lattice  solution
\begin{equation}
\mbox{sn}^2(\phi,\,k)\,=\,\displaystyle\frac{z_1\,z_2}
{z_1\,-\,(z_1\,-\,z_2)\,\mbox{sn}^2(w,\,n)}\,,
\label{latticeI.2}
\end{equation}
with $w\,=\,\sqrt{2\,p\,k^4\,z_1\,(1\,-\,z_2)}\;x$, and the modulus $n$ given by
\begin{equation}
0\,<\,n\,=\,\sqrt{\displaystyle\frac{z_1\,-\,z_2}
{z_1\,(1\,-\,z_2)}}\,\leqslant\,1\,;
\label{n}
\end{equation}
therefore, the soliton lattice has a spatial period $2\,L\,=\,2\,K(n)/
\sqrt{2\,p\,k^4\,z_1\,(1\,-\,z_2)}$. Note that $\phi$ varies between
$\mbox{sn}^{-1}(\sqrt{z_1})$ and $\mbox{sn}^{-1}(\sqrt{z_2})$ [respectively,
$-\,\mbox{sn}^{-1}(\sqrt{z_1})$ and $-\,\mbox{sn}^{-1}(\sqrt{z_2})$] when $x$
varies over one spatial period: indeed, this solution is nontopological. 

In Case I.2(ii) in order to compute the energy for one spatial period $2\,L$, 
it is useful 
to shift the potential by $-\,V^l_{min}\,=\,-\,p\,k^2\,(1\,-\,\Gamma^2)$,
so that the minimum of the potential explored by the soliton lattice is equal to
zero. One finds:
\begin{eqnarray}
&&E_{SL}=4\,\displaystyle\int_{\displaystyle 0}^{\displaystyle L}V_{AL}(\phi)\,dx
\,-\,2\,L\,V^l_{min}\,-\,2\,L\,p^2\,k^2\,a^2\,=\,
2\,\sqrt{\displaystyle\frac{2\,p^2}{z_1\,(1\,-\,z_2)}}\,\nonumber\\
&&\nonumber\\
&&\times\,\left[z_2\,\Pi\left(\displaystyle\frac{z_1\,-\,z_2}{z_1},\,n 
\right)\,-\, \Gamma^2\,
\displaystyle\frac{k^{'2}\,z_2}{1\,-\,k^2\,z_2}\,
\Pi\left(\displaystyle\frac{z_1\,-\,z_2}{z_1\,(1\,-\,k^2\,z_2)},\,n\right)\,-\,
\displaystyle\frac{a^2\,+\,(1\,-\,\Gamma^2)}{2}\right]\,.\nonumber\\
\label{energyI.2}
\end{eqnarray}

The plot of the solutions in Case I.2 is given in Fig.~2(b), 
where the solid  
line corresponds to the nontopological 
soliton lattice  solution of Case I.2(i), 
while the dashed line  corresponds to the nontopological soliton lattice  
solution of Case I.2(ii).
Note that in the appropriate limits (see Sec. IV) these two types 
of solutions give birth to a {\em topological} [Case I.2(i)] 
and a {\em nontopological} 
single soliton [Case I.2(ii)] solution, respectively.

\begin{figure}[htbp]
\begin{center}
\includegraphics[width=0.8\textwidth,angle=-90]{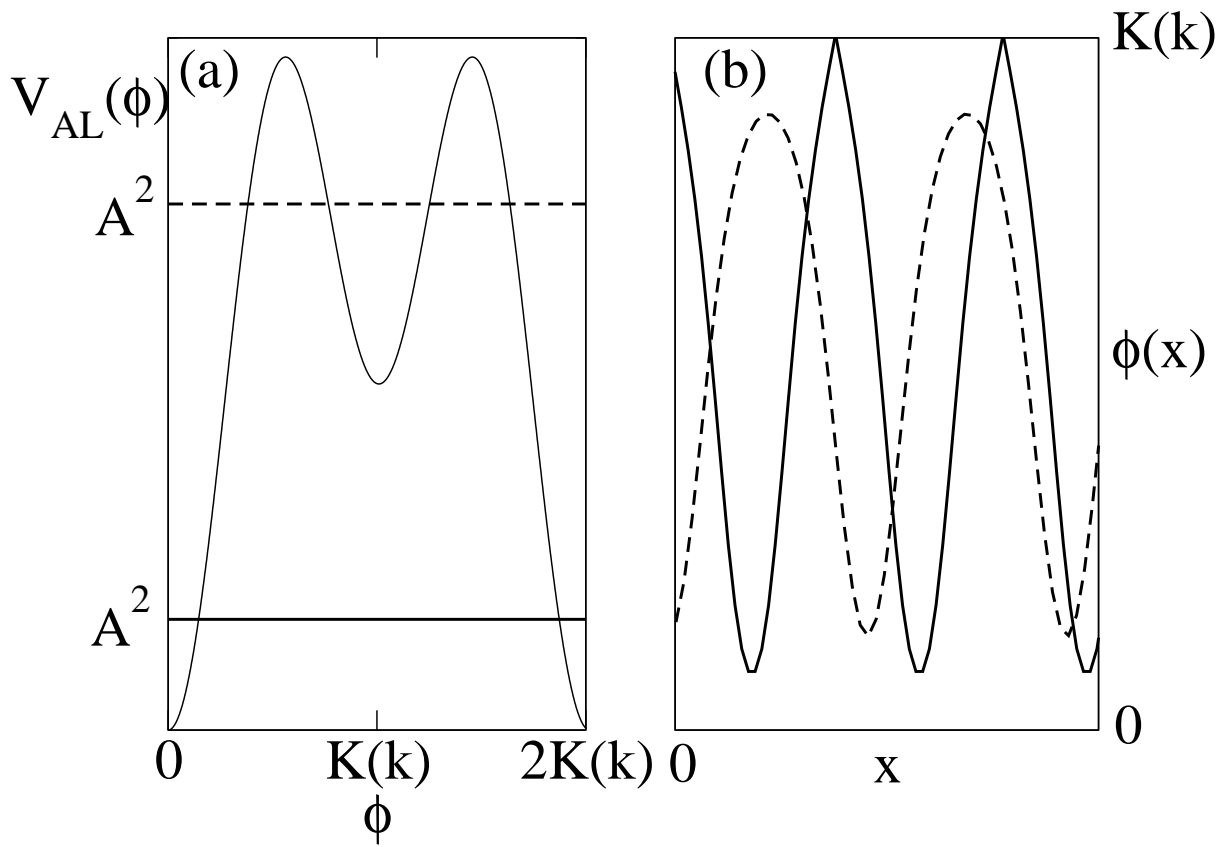}
\end{center}
\caption{Case I.2 : {$k'\,<\,\Gamma\,<\,1$}. 
(a) The shape of the potential $V_{AL}(\phi)$.  The solid and the dashed 
horizontal lines correspond to two choices of the integration parameter 
$A^2$, namely,  Case I.2(i) : {$V_{min}\,\leqslant\,A^2=p\,k^2\,a^2\,<\, 
V^l_{min}$}, i.e., $0\,\leqslant a^2\,<\,1\,-\,\Gamma^2$ (solid line); and
Case I.2(ii) : {$V^l_{min}\,\leqslant\,A^2=p\,k^2\,a^2\,<\,V_{max}$}, i.e.,
$1\,-\,\Gamma^2\,\leqslant\,a^2\,<\,[(1\,-\,\Gamma\,k')/k]^2$ (dashed line).
(b) The soliton lattice solution [i.e., the field $\phi (x)$] with solid 
and dashed curves corresponding to the two values of $A^2$ in (a), i.e., 
respectively, to Case I.2(i) and Case I.2(ii).  The $x$-axis is not labeled 
because the two lattices have different periodicities.} 
\label{fig2}
\end{figure}

\subsection{Case II : $p\,<\,0$ and $q\,\leqslant\,0$ }
\label{caseII}

As explained above, here again it suffices to focus only on the domain 
$0\,\leqslant\,\Gamma\,<\,1$, and we must distinguish between two 
different behaviors of the potential, depending on the value of $\Gamma$.

\subsubsection*{Case II.1 : $0\,\leqslant\,\Gamma\,\leqslant\,k'$}
\label{caseII.1}

The potential $V_{AL}(\phi,\,k)$ has only two extrema in $[0,\,2\,K(k))$, namely an
absolute maximum $V_{max}\,=\,|p|\,k^2\,(1\,-\,\Gamma^2)$ for $\phi\,=\,0$, 
and an absolute minimum $V_{min}\,=\,0$ for $\phi\,=\,K(k)$ [with the choice 
of the shift $C\,=\,|p|\,k^2$ in Eq. (\ref{assocLame})]. A plot of the 
potential is given in Fig.~3(a). 

One can repeat the integration scheme described in the preceding case. 
In particular, considering $z\,=\,\mbox{sn}^2(\phi,\,k)$, Eq. (\ref{quadrature})
becomes:
\begin{equation}
2\,\sqrt{2\,|p|\,k^4}\;x\,=\,\pm\,\int_{\displaystyle 0}^{\displaystyle z}
\displaystyle\frac{d\tilde{z}}{\sqrt{(z_1\,-\,\tilde{z})\,(1\,-\,\tilde{z})\,
(z_2\,-\,\tilde{z})\,(\tilde{z}\,-\,0)}}\,,
\end{equation}
where this time
\begin{equation}
z_{1,\,2}\,=\,\displaystyle\frac{1\,+\,k^2\,-\,\Gamma^2\,-\,a^2\,k^2}{2\,k^2}\,
\left[1\,\pm\,\sqrt{1\,-\,
\displaystyle\frac{4\,k^2\,(1\,-\,\Gamma^2\,-\,a^2)}
{(1\,+\,k^2\,-\,\Gamma^2\,-\,a^2\,k^2)^2}}\;\right]\,,
\label{z12II.1}
\end{equation} 
$z_1\,>\,1\,\geqslant\,z_2\,>\,z\,>\,0$, and $0\leqslant a^2<(1-\Gamma^2)$. This integral can be evaluated using
formula 3.147(2) of \cite{gradshteyn} 
and one obtains the following {\em nontopological} soliton lattice:
\begin{equation}
\mbox{sn}^2(\phi,\,k)\,=\,\displaystyle\frac{z_1\,z_2\,\mbox{sn}^2(y,\,r)}
{(z_1\,-\,z_2)\,+\,z_2\,\mbox{sn}^2(y,\,r)}\;.
\label{latticeII.1}
\end{equation}
Here $y\,=\,\sqrt{2\,|p|\,k^4\,(z_1\,-\,z_2)}\;x$ and the modulus $r$ that 
controls the density of solitons in the lattice is given by
\begin{equation}
0\,<\,r\,=\,\displaystyle\sqrt{\frac{z_2\,(z_1\,-\,1)}{z_1\,-\,z_2}}\,
\leqslant\,1\,.
\label{r}
\end{equation}
The period of the lattice is $2\,L\,=\,2\,K(r)/\sqrt{2\,|p|\,k^4\,
(z_1\,-\,z_2)}$ and one notices that $\phi$ varies between $0$ and 
$\mbox{sn}^{-1}(\sqrt{z_2})$ [respectively, $0$ and 
$-\,\mbox{sn}^{-1}(\sqrt{z_2})$ ] in one period.
A plot of the solution is given in Fig.~3(b).

\begin{figure}[htbp]
\begin{center}
\includegraphics[width=0.8\textwidth,angle=-90]{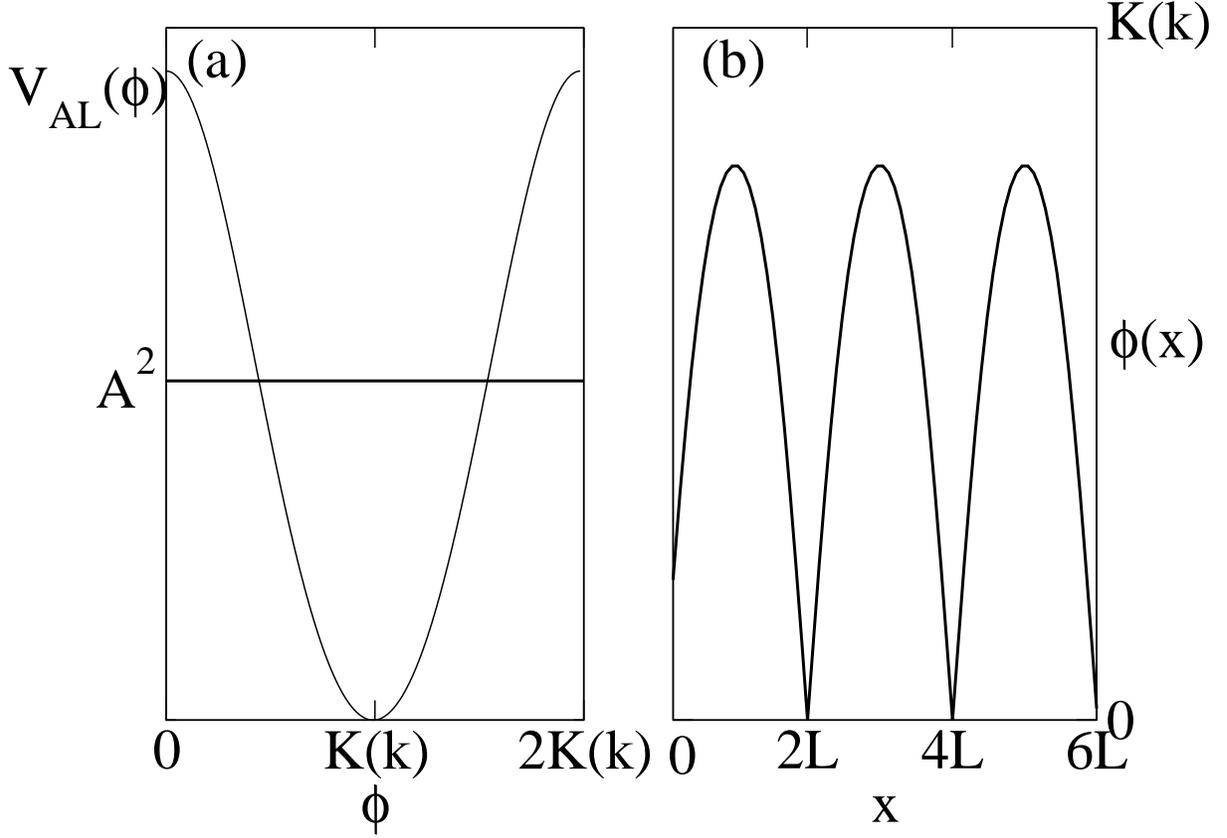}
\end{center}
\caption{Case II.1 : $0\,\leqslant\,\Gamma\,\leqslant\,k'$.
(a) The shape of the potential $V_{AL}(\phi)$.  The 
solid horizontal line corresponds 
to  the integration parameter $A^2$.
(b) The soliton lattice solution [i.e., the field $\phi (x)$] 
corresponding to the value of $A^2$ in (a).} 
\label{fig3}
\end{figure}

One can compute the energy for one period $2\,L$ of this lattice,
\begin{eqnarray}
E_{SL}&=&2\,\sqrt{\displaystyle\frac{2\,|p|}{z_1\,-\,z_2}}\,\left\{
\left[-(z_1\,-\,1)\,K(r)\,+\,z_1\,\Pi\left(-\,
\displaystyle\frac{z_2}{z_1\,-\,z_2},\,r\right)\right]\,\right.\nonumber\\
&&\nonumber\\
&&\left.-\,\Gamma^2\,\left[-\,\displaystyle\frac{z_1\,-\,1}
{1\,-\,z_1\,k^2}\,K(r)\,+\,\displaystyle\frac{z_1\,k^{'2}}{1\,-\,z_1\,k^2}
\,\Pi\left(-\,\displaystyle\frac{z_2\,(1\,-\,z_1\,k^2)}{z_1\,-\,z_2},\,r\right)
\right]\,-\,\displaystyle\frac{a^2\,K(r)}{2}\right\}\,.\nonumber\\
\label{energyII.1}
\end{eqnarray}

\subsubsection*{Case II.2 : $k'\,<\,\Gamma\,<\,1$}
\label{caseII.2}

In this case, the potential $V_{AL}(\phi,\,k)$ has two maxima and two 
degenerate minima in one period $0\,\leqslant\,\phi\,<\,2\,K(k)$. Namely, 
an absolute maximum, 
$V_{max}\,=\,|p|\,(1\,-\,\Gamma\,k')^2$ (for $\phi\,=\,0$);
a relative maximum $V^l_{max}\,=\,|p|\,(\Gamma\,-\,k')^2$ [for $\phi\,=\,K(k)$]:
and two  absolute minima $V_{min}\,=\,0$ situated symmetrically around $K(k)$
[for
$\phi\,=\,\mbox{sn}^{-1}\sqrt{(1\,-\,\Gamma\,k')/k^2},\,
2\,K(k)\,-\,\mbox{sn}^{-1}\sqrt{(1\,-\,\Gamma\,k')/k^2}]$.
Note that we used a shift $C\,=\,|p|\,(1\,-\,2\,\Gamma\,k'\,+\Gamma^2)$ 
in the expression (\ref{assocLame}) of the potential.
A plot of the potential as a function of $\phi$ is given in Fig.~4(a). 

The very existence of adjacent degenerate minima of the potential 
separated by different barriers  to the right and to the left
(on the $\phi$ axis) implies that in this case one will have 
two different soliton lattices (and, correspondingly, two different 
topological single soliton solutions).  This is similar to the 
DSG case \cite{leung,peyrard,dando}.

We follow the same type of integration procedure as that described
above. 

{\em Case II.2(i)} :  When
{$V_{min}\,\leqslant\,A^2=|p|\,k^2\,a^2\,<\,V^l_{max}$}, i.e., 
$0\,\leqslant\,a^2\,<\,(\Gamma\,-\,k')^2/k^2$,
Eq.~(\ref{quadrature}), under the change of variable
$z\,=\,\mbox{sn}^2(\phi,\,k)$, leads to two different types of solutions.

{\em Solution 1} : for $0\,<\,z\,\leqslant\,z_2\,<\,z_1\,<\,1$, 
one has:
\begin{equation}
2\,\sqrt{2\,|p|\,k^4}\;x\,=\,\pm
\displaystyle\int_{\displaystyle 0}^{\displaystyle z}
\displaystyle\frac{d\tilde{z}}{\sqrt{(1\,-\,\tilde{z})\,(z_1\,-\,\tilde{z})\,
(z_2\,-\,\tilde{z})\,(\tilde{z}\,-\,0)}}\,,
\label{caseII2i1}
\end{equation}
where now:
\begin{equation}
z_{1,\,2}\,=\,\displaystyle\frac{2\,(1\,-\,\Gamma\,k')\,-\,a^2\,k^2}{2\,k^2}\,
\left\{1\,\pm\,\sqrt{1\,-\,
\displaystyle\frac{4\,[(1\,-\,\Gamma\,k')^2\,-\,a^2\,k^2]}
{[2\,(1\,-\,\Gamma\,k')\,-\,a^2\,k^2]^2}}\;\right\}\,.
\label{z12II.2}
\end{equation}
Equation (\ref{caseII2i1}) can be integrated using formula 3.147(2) of
\cite{gradshteyn}, thus
obtaining:
\begin{equation}
\mbox{sn}^2(\phi,\,k)\,=\,\displaystyle\frac{z_2\,\mbox{sn}^2(w,\,s)}
{(1\,-\,z_2)\,+\,z_2\,\mbox{sn}^2(w,\,s)}\,,
\label{latticeII.2.1}
\end{equation}
with $w\,=\,\sqrt{2\,|p|\,k^4\,z_1\,(1\,-\,z_2)}\;x$ and the modulus
\begin{equation}
0\,<\,s\,=\,\sqrt{\displaystyle\frac{(1\,-\,z_1)\,z_2}{(1\,-\,z_2)\,z_1}}\,
\leqslant\,1\,.
\label{s}
\end{equation}
It is a {\em nontopological} soliton lattice, 
with $\phi$ oscillating between $0$
and $\mbox{sn}^{-1}(\sqrt{z_2})$ [respectively, $0$ and 
$-\,\mbox{sn}^{-1}(\sqrt{z_2})$]. Its energy for one period $2\,L\,=\,2\,K(s)/
\sqrt{2\,|p|\,k^4\,z_1\,(1\,-\,z_2)}$ is given by:
\begin{eqnarray}
E_{SL}\,=\,2\,\sqrt{\displaystyle\frac{2\,|p|}{z_1\,(1\,-\,z_2)}}\left\{
\left[\displaystyle\frac{(\Gamma\,-\,k')^2}{k^2}\,-\,
\displaystyle\frac{a^2}{2}\right]\,K(s)\right.&+&
\Pi\left(-\,\displaystyle\frac{z_2}{1\,-\,z_2},\,s\right)\nonumber\\
&-&\left.\Gamma^2\,
\Pi\left(-\,\displaystyle\frac{z_2\,k^{'2}}{1\,-\,z_2},\,s\right)\right\}\,.
\label{energyII.2.1}
\end{eqnarray}

{\em Solution 2} : for $0\,<\,z_2\,<\,z_1\,\leqslant\,z\,<\,1$ one obtains
from Eq. (\ref{quadrature}):
\begin{equation}
2\,\sqrt{2\,|p|\,k^4}\;x\,=\,\pm\,
\displaystyle\int_{\displaystyle z}^{\displaystyle 1}
\displaystyle\frac{d\tilde{z}}
{\sqrt{(1\,-\,\tilde{z})\,(\tilde{z}\,-\,z_1)\,
(\tilde{z}\,-\,z_2)\,(\tilde{z}\,-\,0)}}\,,
\label{caseII2i2}
\end{equation}
with $z\,=\,\mbox{sn}^2(\phi,\,k)$ and $z_{1,\,2}$ given by Eq. (\ref{z12II.2}).
This integral can be solved using formula 3.147(7) of \cite{gradshteyn}, thus
leading to the following {\em nontopological} pulse-like lattice
\begin{equation}
\mbox{sn}^2(\phi,\,k)\,=\,\displaystyle\frac{z_1}
{z_1\,+\,(1\,-\,z_1)\,\mbox{sn}^2(w,\,s)}\,,
\label{latticeII.2.2}
\end{equation}
with $w\,=\,\sqrt{2\,|p|\,k^4\,z_1\,(1\,-\,z_2)}\;x$ and the modulus $s$ given by
the same expression as above, Eq. (\ref{s}). Of course, one notices immediately
that this soliton lattice is different from the previous one, Eq. 
(\ref{latticeII.2.1}). In one period $2\,L\,=\,2\,K(s)/
\sqrt{2\,|p|\,k^4\,z_1\,(1\,-\,z_2)}$,  $\phi$ oscillates between 
$\mbox{sn}^{-1}\sqrt{z_1}$ and $K(k)$ [respectively, $-\,\mbox{sn}^{-1}\sqrt{z_1}$ 
and $-K(k)$]. 

The energy corresponding to one period $2\,L$ of this pulse lattice is
given by:
\begin{eqnarray}
E_{SL}\,=\,2\,\sqrt{\displaystyle\frac{2\,|p|}{z_1\,(1\,-\,z_2)}}\left\{
\left[\displaystyle\frac{(1\,-\,\Gamma\,k')^2}{k^2}\,-\,
\displaystyle\frac{a^2}{2}\right]\,K(s)\right.&+&
\Pi\left(-\,\displaystyle\frac{1\,-\,z_1}{z_1},\,s\right)\nonumber\\
&+&\left.\Gamma^2\,
\Pi\left(-\,\displaystyle\frac{1\,-\,z_1}{z_1\,k^{'2}},\,s\right)\right\}\,.
\label{energyII.2.2}
\end{eqnarray}

A plot of the solutions is given in Fig.~4(b)
where the solid and dashed lines correspond, respectively,  
to the Solution 1 and Solution 2. 
Therefore, as expected, we obtained two different soliton lattices 
(with two different limiting cases of topological single solitons). 
Note that the terms ``large" and
``small" applied to the topological solitons do not refer, as usual, to the 
length of the interval they cover in $\phi$ space, but to the height of the
barrier they ``encounter" \cite{peyrard}.

\begin{figure}[htbp]
\begin{center}
\includegraphics[width=0.8\textwidth,angle=-90]{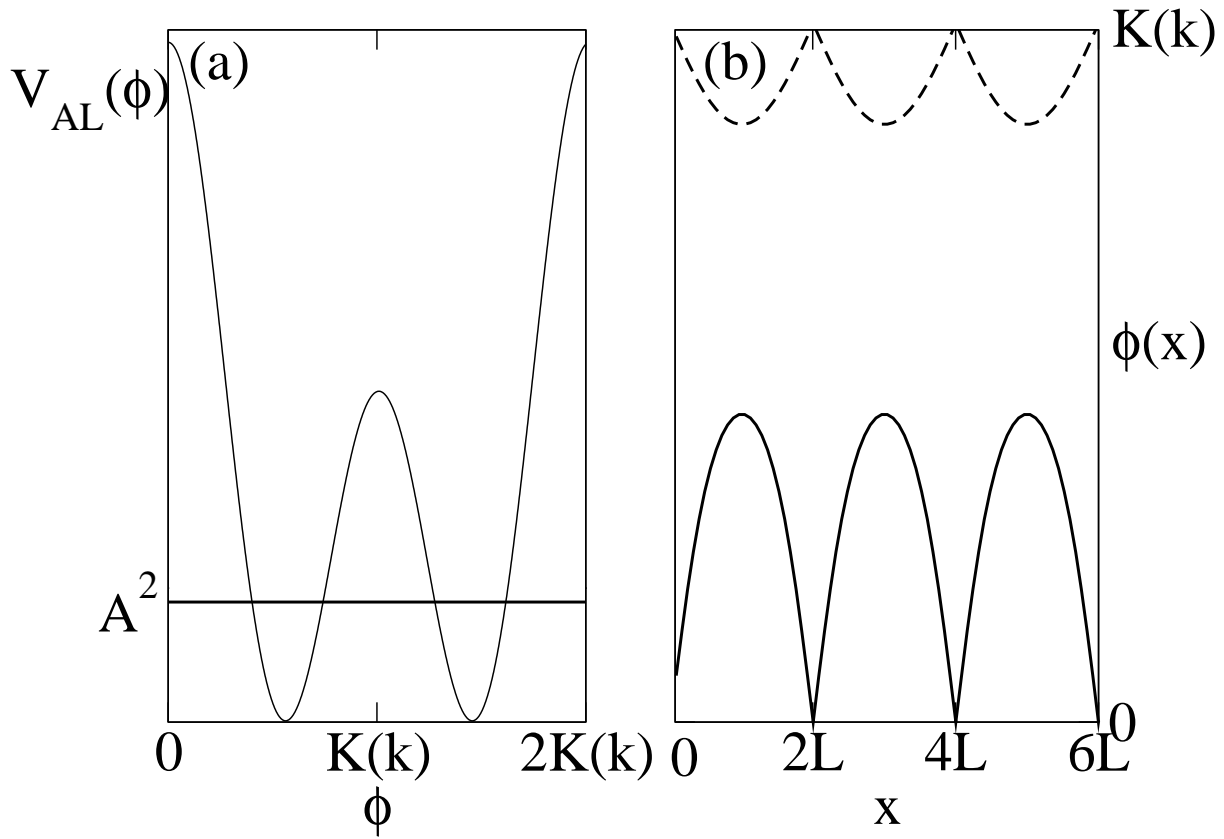}
\end{center}
\caption{Case II.2 : $k'\,<\,\Gamma\,<\,1$.
(a) The shape of the potential $V_{AL}(\phi)$.  The 
solid horizontal line corresponds 
to  the integration parameter $A^2$.
(b) The two nontopological soliton lattice solutions [i.e., the field 
$\phi (x)$] corresponding to the value of $A^2$ in (a); the solid line 
represents Solution 1 (see the main text), while the dashed 
line--Solution 2.} 
\label{fig4}
\end{figure}

{\em Case II.2(ii)} :  When {$V^l_{max}\,\leqslant\,A^2=|p|\,k^2\,a^2\,<\, 
V_{max}$}, i.e., 
$(\Gamma\,-\,k')^2/k^2\,\leqslant\,a^2\,<\,(1\,-\,\Gamma\,k')^2/k^2$
there is no soliton lattice solution.

\subsection{Case III : $p\,>\,0$ and $q\,\leqslant \,0$}
\label{caseIII}
Unlike the last two cases, here $p^2$ can be $>$ or $<$ $q^2$. However, it 
turns out that
in this case, whatever the value of $\Gamma\,=\,\sqrt{|q|/p}$, the potential 
$V_{AL}(\phi)$ has
only two extrema in one period $0\,\leqslant\,\phi\,<\,2\,K(k)$, namely an
absolute minimum $V_{min}\,=\,0$ for $\phi\,=\,0$ and a maximum  
$V_{max}\,=\,p\,k^2\,(1\,+\,\Gamma^2)$ for $\phi\,=\,K(k)$. Note that we 
chose $C=p\,k^2\,\Gamma^2$ in Eq. (\ref{assocLame}). 
A plot of the potential $V_{AL}(\phi)$ is given in Fig.~5(a).  With
$z\,=\,\mbox{sn}^2(\phi,\,k)$, Eq. (\ref{quadrature}) now becomes:
\begin{equation}
2\,\sqrt{2\,p\,k^4}\;x\,=\,\pm\,
\displaystyle\int_{\displaystyle z}^{\displaystyle 1}
\displaystyle\frac{d\tilde{z}}
{\sqrt{(z_1\,-\,\tilde{z})\,(1\,-\,\tilde{z})\,
(\tilde{z}\,-\,z_2)\,(\tilde{z}\,-\,0)}}\,,
\label{caseIII.}
\end{equation}
with 
\begin{equation}
z_{1,\,2}\,=\,\displaystyle\frac{(1\,+\,\Gamma^2\,k^{'2}\,+\,a^2\,k^2)}{2\,k^2}\,
\left[1\,\pm\sqrt{1\,-\,\displaystyle\frac{4\,k^2\,a^2}
{(1\,+\,\Gamma^2\,k^{'2}\,+\,a^2\,k^2)^2}}\;\right]\,,
\label{z12III}
\end{equation}
$z_1\,>\,1\,\geqslant\,z\,>\,z_2\,>\,0$. Using  the formula 3.147(5) 
of \cite{gradshteyn}, one can
evaluate the integral in Eq. (\ref{caseIII.}), thus obtaining a 
{\em nontopological} soliton lattice 
\begin{equation}
\mbox{sn}^2(\phi,\,k)\,=\,\displaystyle
\frac{[(z_1\,-\,z_2)\,-\,z_1\,(1\,-\,z_2)\,\mbox{sn}^2(y,\,t)]}
{[(z_1\,-\,z_2)\,-\,(1\,-\,z_2)\,\mbox{sn}^2(y,\,t)]}\,.
\label{latticeIII}
\end{equation}
Here $y\,=\,\sqrt{2\,p\,k^4\,(z_1\,-\,z_2)}\;x$ and the elliptic modulus:
\begin{equation}
0\,<\,t\,=\,\sqrt{\displaystyle\frac{z_1\,(1\,-\,z_2)}{z_1\,-\,z_2}}\,\leqslant\,1\,.
\end{equation}
In one period $2\,L\,=\,2\,K(t)/\sqrt{2\,p\,k^4\,(z_1\,-\,z_2)}$, $\phi$
oscillates between $\mbox{sn}^{-1}\sqrt{z_2}$ and $K(k)$ [respectively,
$-K(k)$ and $-\,\mbox{sn}^{-1}\sqrt{z_2}$].
A plot of the solution is given in Fig.~5(b).

\begin{figure}[htbp]
\begin{center}
\includegraphics[width=0.8\textwidth,angle=-90]{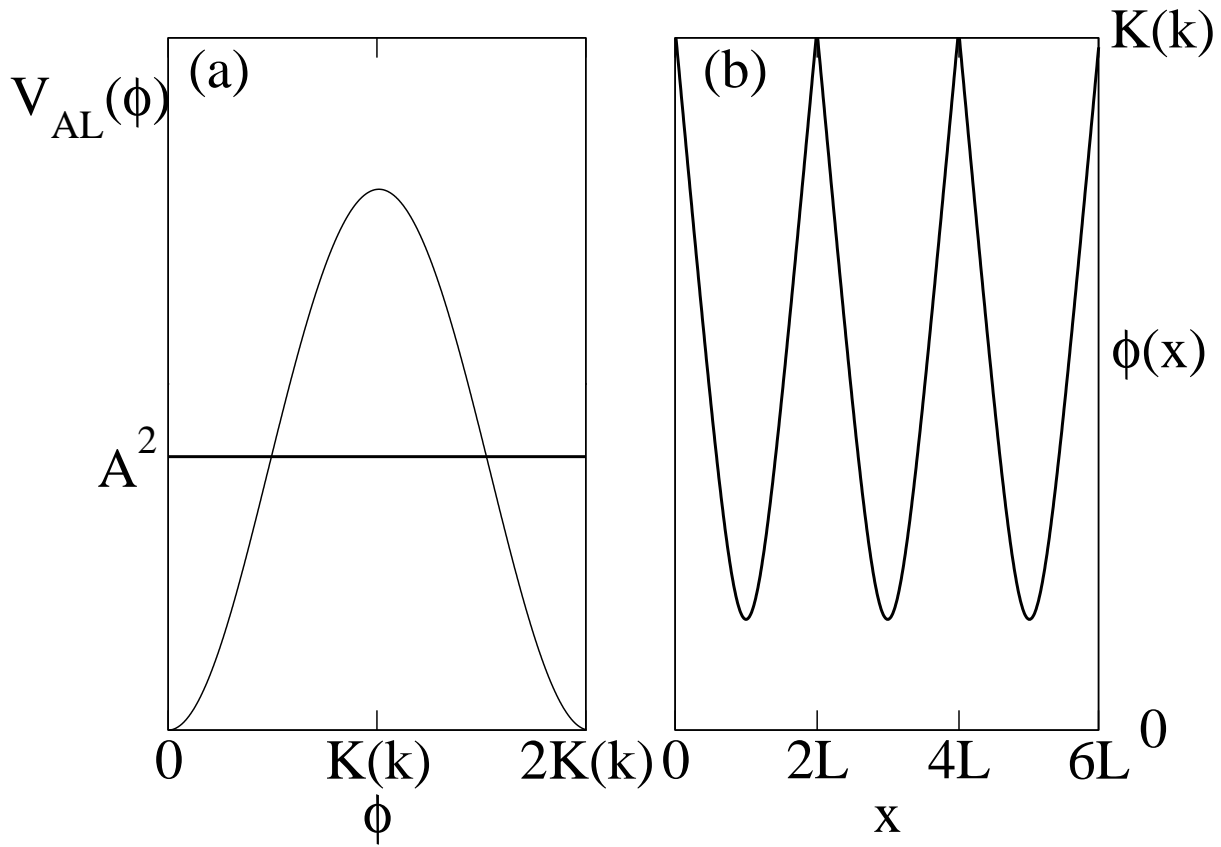}
\end{center}
\caption{Case III : $p\,>\,0$ and $q\,\leqslant \,0$.
(a) The shape of the potential $V_{AL}(\phi)$.  The 
solid horizontal line corresponds 
to  the integration parameter $A^2$.
(b) The soliton lattice solution [i.e., the field $\phi (x)$] 
corresponding to the value of $A^2$ in (a).} 
\label{fig5}
\end{figure}

The energy corresponding to one period $2\,L$ of this lattice is:
\begin{eqnarray}
E_{SL}&=&2\,\sqrt{\displaystyle\frac{2\,p}{z_1\,-\,z_2}}\,\left\{
\left[z_1\,K(t)\,-\,(z_1\,-\,1)\,
\Pi\left(\displaystyle\frac{1\,-\,z_2}{z_1\,-\,z_2},\,t\right)\right]\,-\,\right.
\Gamma^2\,k^{'2}\,\left[\displaystyle\frac{z_1}{z_1\,k^2\,-\,1}K(t)\,-\,\right.
\nonumber\\
&&\nonumber\\
&&\left.\left.-\,
\displaystyle\frac{z_1\,-\,1}{k^{'2}\,(z_1\,k^2\,-\,1)}\,
\Pi\left(-\,\displaystyle\frac{(1\,-\,z_2)\,(z_1\,k^2\,-\,1)}
{k^{'2}\,(z_1\,-\,z_2)},\,t\right)\right]\,-\,
\displaystyle\frac{a^2\,K(t)}{2}\right\}\,.
\label{energyIII}
\end{eqnarray}

\section{Single Soliton Solutions of The AL Potential}
Having obtained the soliton lattice solutions of the AL problem, it is 
now straightforward to consider the specific limits of the integration constant $A^2$ 
and to obtain the 
corresponding {\em single soliton} solutions of the Lam\'e potential and compute 
their energy. 

{\bf Cases I.1 and I.2(i)} : The limit of a  single soliton 
(that corresponds to the period of the lattice
$L\,\rightarrow\,\infty$)  is obtained for $t\,\nearrow 1$ [$a^2\,\searrow\,0$;
$z_1\,\rightarrow (1\,-\,\Gamma^2\,k^{'2})/k^2$ and $z_2\,\searrow\,0$].
One notices that now
$\phi$ can vary continuously  between $-K(k)$ and $K(k)$ while $x$ explores the
whole real axis; therefore, one obtains a {\em topological single-soliton}
(kink-like) solution,
\begin{equation}
{\mbox{sn}(\phi,\,k)}\,=\,\pm\displaystyle
\sqrt{\frac{1\,-\,\Gamma^2 \,k^{'2}}
{1\,-\,\Gamma^2\,k^{'2}+k'^2(1-\Gamma^2){\mbox{sinh}^2(y^*)}}}\;\, ,
\label{singleI.1}
\end{equation}
where $y^*\,=\,\sqrt{2\,p\,k^2\,(1\,-\,\Gamma^2\,k^{'2})}\;x$.
The corresponding energy of this single kink-like soliton is obtained 
from Eq. (\ref{energyI.1}), using the relations 
\begin{equation}
\mbox{lim}_{r \nearrow 1} \Pi(n^2,r)=\mbox{lim}_{r \nearrow 1} \frac{K(r)}{1-n^2}
-\frac{n}{2(1-n^2)}\,\mbox{ln}\left(\frac{1+n}{1-n}\right)\,.
\label{limit}
\end{equation}

It may be noted that all the divergences cancel mutually, thus leading to
a finite energy of the single soliton:  
\begin{equation}
E_K\,=\,\sqrt{2\,p\,k^2}\left[\displaystyle\frac{1}{k}\,\mbox{ln}\left(
\displaystyle\frac{\sqrt{1\,-\,\Gamma^2\,k^{'2}}\,+\,k}
{\sqrt{1\,-\,\Gamma^2\,k^{'2}}\,-\,k}\right)\,-\,\Gamma^2\,
\displaystyle\frac{1}{(\Gamma\,k)}\,\mbox{ln}\left(
\displaystyle\frac{\sqrt{1\,-\,\Gamma^2\,k^{'2}}\,+\,(\Gamma\,k)}
{\sqrt{1\,-\,\Gamma^2\,k^{'2}}\,-\,(\Gamma\,k)}\right)\right]\,.
\label{energysingleI.1}
\end{equation}

Following Manton \cite{manton} one can compute 
the {\em asymptotic} interaction energy between two 
solitons in the array, or between a soliton and an anti-soliton. 
As an illustration we will consider this particular soliton solution; 
for all the other cases 
one can follow exactly the same procedure. 

From Eq. (\ref{singleI.1}) one obtains
the asymptotic shape of the soliton, e.g., for $x \rightarrow + \infty $:
\begin{equation}
\phi_{as}\approx \left[\frac{4(1-\Gamma ^2\,k'^2)}{k'^2\,(1-\Gamma ^2)}\right]^{1/2} \,
\mbox{exp}\left(-\sqrt{2\,p\,k^2 \,(1-\Gamma ^2\,k'^2)} x\right)\,.
\end{equation}
Then, if $2\,L\,\gg 1 $ is the distance between two solitons in the array, 
according to \cite{manton} the asymptotic interaction energy is
\begin{equation}
U(2\,L)\approx - \frac{8\,(2\,p\,k^2~)^{1/2}\,(1-\Gamma ^2\,k'^2)^{3/2}}{k'^2\,(1-\Gamma ^2)}\,
\mbox{exp}\left(-2\,L\,\sqrt{2\,p\,k^2 \,(1-\Gamma ^2\,k'^2)}\right)\,.
\end{equation}
If one considers now the small parameter $ a^2\,\ll \,1$ that measures the distance from the single soliton limit,
according to the results in Sec. III we can obtain the asymptotic expression 
for $2\,L$ as:
\begin{equation}
2\,L\approx\frac{1}{\sqrt{2\,p\,k^2 \,(1-\Gamma ^2\,k'^2)}}\,\mbox{ln}
\left(\frac{16\,(1-\Gamma ^2\,k'^2)^2}{a^2\,k'^2\,(1-\Gamma^2)}\right)\,,
\end{equation}
and thus
\begin{equation}
U(a^2)\approx \sqrt{\frac{p\,k^2}{2\,(1-\Gamma^2\,k'^2)}}\,a^2\,,
\end{equation}
that corresponds to a {\em repulsive asymptotic} interaction between 
two solitons in the array.  One can consider also the {\em asymptotic} 
interaction energy between a soliton and an anti-soliton and obtains 
simply $U(a^2)\approx -\sqrt{\frac{p\,k^2}{2\,(1-\Gamma^2\,k'^2)}}\,a^2$, 
i.e., an {\em attractive} interaction.  

\subsubsection*{Sine-Gordon Limit}
Taking now the limit
$k\,\rightarrow\,0$, with $|p|,\,|q|\,\rightarrow \infty$, so that 
$|p|\,k^2\,\rightarrow\, P$ = finite and $|q|\,k^2\,\rightarrow\, Q$ = finite,
in the expressions of
the AL single soliton solution, Eq. (\ref{singleI.1}),
one obtains the correct kink-like solution of the
{sine-Gordon potential} in Eq. (\ref{sG}), namely
\begin{equation}
\sin\phi_{sG}\,=\,\pm\mbox{sech}(y^*)\,,
\label{sGkinkI.1}
\end{equation}
with $y^*\,=\,\sqrt{2\,P\,(1\,-\,\Gamma^2)}\;x$, and the energy
\begin{equation}
E_{sG}\,=\,2\,\sqrt{2\,P\,(1\,-\,\Gamma^2)}\,.
\label{sGenergyI.1}
\end{equation}

{\bf Case I.2(ii)} : The single soliton  limit $L\,\rightarrow\,\infty$
[i.e., $n\,\nearrow\,1$; 
$a^2\,\searrow\,(1\,-\,\Gamma^2)$, $z_1\,\rightarrow 1$, and
$z_2\,\rightarrow\,(1\,-\,\Gamma^2)/k^2$] reads:
\begin{equation}
\displaystyle{\mbox{sn}(\phi,\,k)}\,=\,
\pm\,\sqrt{\frac{1-\Gamma^2}{1\,-\,\Gamma^2
+(\Gamma^2 -k'^2){\mbox{sech}^2(\omega^*)}}}\,\; , 
\label{singleI.2}
\end{equation}
with $\omega^*\,=\,\sqrt{2\,p\,k^2\,(\Gamma^2\,-\,k^{'2})}\;x$. 
It represents a {\em nontopological} soliton, with $\phi$ varying
between $K(k)$ and $\mbox{sn}^{-1}(\sqrt{(1\,-\,\Gamma^2)/k^2})$
[respectively,  $-\,K(k)$ and $-\,\mbox{sn}^{-1}(\sqrt{(1\,-\,\Gamma^2)/k^2})$]
when $x$ runs over the real axis, i.e., it does not connect two 
adjacent degenerate minima of the potential.

Its energy is obtained from Eq. (\ref{energyI.2}) and reads:
\begin{equation}
E_{S}\,=\,\sqrt{2\,p\,k^2}
\left[\Gamma^2\,\displaystyle\frac{1}{(\Gamma\,k)}\,
\mbox{ln}\left(\displaystyle\frac{(\Gamma\,k)\,+\,\sqrt{\Gamma^2\,-\,k^{'2}}}
{(\Gamma\,k)\,-\,\sqrt{\Gamma^2\,-\,k^{'2}}}\right)\,-\,
\displaystyle\frac{1}{k}\,
\mbox{ln}\left(\displaystyle\frac{k\,+\,\sqrt{\Gamma^2\,-\,k^{'2}}}
{k\,-\,\sqrt{\Gamma^2\,-\,k^{'2}}}\right)\right].
\label{energysingleI.2}
\end{equation}

{\bf Case II.1} : Consider now the {single soliton} limit $r\,\nearrow \,1$ 
[$a^2\,\searrow 0$, $z_1\,\rightarrow \,(1\,-\,\Gamma^2)/k^2$, and 
$z_2\,\rightarrow 1$]. One obtains the following {\em topological} (i.e.,
kink-like) soliton:
\begin{equation}
\displaystyle{\mbox{sn}(\phi,\,k)}\,=\,
\pm\,\sqrt{\displaystyle\frac{1\,-\,\Gamma^2}
{k^{'2}\,-\,\Gamma^2+(1-\Gamma^2)\mbox{sinh}^2(y^*)}}\;
\mbox{sinh}(y^*)\,,
\label{singleII.1}
\end{equation}
with $y^*\,=\,\sqrt{2\,|p|\,k^2\,(k^{'2}\,-\,\Gamma^2)}\;x$, of energy
\begin{equation}
E_{K}\,=\,2\,\sqrt{2\,|p|\,k^2}\;
\left[\displaystyle\frac{1}{k}\,
\mbox{arctan}\left(\displaystyle\frac{k}{\sqrt{k^{'2}\,-\,\Gamma^2}}\right)\,-\,
\Gamma^2\,\displaystyle\frac{1}{(\Gamma\,k)}\,
\mbox{arctan}\left(\displaystyle\frac{(\Gamma\,k)}
{\sqrt{k^{'2}\,-\,\Gamma^2}}\right)\right] , 
\label{energysingleII.1}
\end{equation}
where we took the limit $r\,\nearrow\,1$ in Eq. (\ref{energyII.1}) 
using the relation 
\begin{equation}
\mbox{lim}_{r \nearrow 1} \Pi(-n^2,r)=\mbox{lim}_{r \nearrow 1} \frac{K(r)}{1+n^2}
+\frac{n}{1+n^2} \arctan(n)\,.
\label{limitagain}
\end{equation}

Note that one can also consider the limit of the sine-Gordon potential 
in this case and obtain the correct sine-Gordon soliton.

{\bf Case II.2(i) : Solution 1 :  Large Topological Kink} : 
Considering the single soliton limit, $s\,\nearrow\,1$ 
[$a^2\,\searrow\,0$, $z_{1,\,2}\,\rightarrow\,(1\,-\,\Gamma\,k')/k^2$],
one obtains a {\em ``large" topological kink}, that interpolates between two
adjacent minima ``across" the large barrier $V_{max}$ of the potential:
\begin{equation}
\displaystyle{\mbox{sn}(\phi,\,k)}
\,=\,\pm\,
\sqrt{\displaystyle\frac{1\,-\,\Gamma\,k'}
{k'\,(\Gamma\,-\,k')+(1-\Gamma k')\mbox{tanh}^2(\omega^*)}}\;
\mbox{tanh}(\omega^*)\,,
\label{singleII.2.1}
\end{equation}
with $\omega^*\,=\,\sqrt{2\,|p|\,k'\,(1\,-\,\Gamma\,k')\,(\Gamma\,-\,k')}\;x$.
Its energy is found from Eqs. (\ref{energyII.2.1}) and (\ref{limitagain}) as:
\begin{equation}
E_{K}\,=\,2\,\sqrt{2\,|p|}\,
\left[\mbox{arctan}\sqrt{\displaystyle\frac{1\,-\,\Gamma\,k'}
{k'\,(\Gamma\,-\,k')}}\,-\,\Gamma\,
\mbox{arctan}\sqrt{\displaystyle\frac{k'\,(1\,-\,\Gamma\,k')}
{\Gamma\,-\,k'}}\;\right]\,.
\label{energysingleII.2.1}
\end{equation}

{\bf Case II.2(i) : Solution 2 : Small Topological Kink} :
Considering now the {\em single soliton} limit $s\,\nearrow\,1$ 
[$a^2\,\rightarrow\,0$, $z_{1,\,2}\,\rightarrow\,(1\,-\,\Gamma\,k')/k^2$],
one obtains a {\em ``small" topological kink}, that interpolates between two
adjacent minima ``across" the small barrier $V'_{max}$ of the potential:
\begin{equation}
\displaystyle
{\mbox{sn}(\phi,\,k)}\,=\,\pm\,
\sqrt{\displaystyle\frac{1-\Gamma\,k'}
{1\,-\,\Gamma\,k'+k'(\Gamma-k')\mbox{tanh}^2(\omega^*)}}\,,
\label{singleII.2.11}
\end{equation}
with $\omega^*\,=\,\sqrt{2\,|p|\,k'\,(1\,-\,\Gamma\,k')\,(\Gamma\,-\,k')}\;x$.
Its energy is found from Eqs. (\ref{energyII.2.2}) and (\ref{limitagain}):
\begin{equation}
E_{K}\,=\,2\,\sqrt{2\,|p|}\,
\left[\Gamma\,\mbox{arctan}\sqrt{\displaystyle\frac{\Gamma\,-\,k'}
{k'\,(1\,-\,\Gamma\,k')}}\,-\,
\mbox{arctan}\sqrt{\displaystyle\frac{k'\,(\Gamma\,-\,k')}
{1\,-\,\Gamma\,k'}}\;\right]\,.
\label{energysingleII.2.2}
\end{equation}

{\bf Case III} : 
The {single soliton} limit $t\,\nearrow\,1$ [$a^2\,\searrow \,0$,
$z_1\,\rightarrow\,(1\,+\,\Gamma^2\,k^{'2})/k^2$, and $z_2\,\rightarrow\,0$]
represents a {\em topological soliton}
\begin{equation}
\displaystyle{\mbox{sn}(\phi,\,k)}
\,=\,\pm\,\sqrt{\displaystyle\frac{(1\,+k'^2\,\Gamma^2)}
{1\,+\,\Gamma^2\,k^{'2}+k'^2(1+\Gamma^2)\mbox{sinh}^2(y^{*})}}\; , 
\label{singleIII}
\end{equation}
with $y^*\,=\,\sqrt{2\,p\,k^2\,(1\,+\,\Gamma^2\,k^{'2})}\;x$.
Its energy is given by:
\begin{equation}
E_K\,=\,\sqrt{2\,p\,k^2}\,\left[
\displaystyle\frac{1}{k}\,\mbox{ln}
\left(\displaystyle\frac{\sqrt{1\,+\,\Gamma^2\,k^{'2}}\,+\,k}
{\sqrt{1\,+\,\Gamma^2\,k^{'2}}\,-\,k}\right)\,+\,
2\,\Gamma^2\,
\displaystyle\frac{1}{(\Gamma\,k)}\,\mbox{arctan}
\left(\displaystyle\frac{(\Gamma\,k)}
{\sqrt{1\,+\,\Gamma^2\,k^{'2}}}\right)\right]\,.
\label{energysingleIII}
\end{equation}

Here again one can consider the limit of the sine-Gordon potential and 
obtain the correct sine-Gordon  soliton of energy $E_{sG}=2\,\sqrt{2\,P\,(1 
+\Gamma^2)}$ .

\section{Soliton Lattice and Single Soliton Solutions of the Lam\'e Potential}
We shall now demonstrate that having obtained the AL soliton lattice and single soliton solutions, one can 
immediately obtain the corresponding solutions of the Lam\'e problem
by taking suitable limits of the corresponding AL solutions.

\subsection{Lam\'e Soliton Lattice Solutions}
Starting from the AL soliton lattice solutions and considering
the limit $\Gamma \rightarrow 0$, we obtain the two following  types of Lam\'e soliton lattice
solutions:

{\bf Cases I.1 and III ($p>0$)} : These lead to  {\em type I} Lam\'e soliton lattice solution.
The solution is  given by Eq. (\ref{latticeI.1}) with 
simpler forms for  $z_1,z_2$, namely $z_1={1}/{k^2}$, $z_2=a^2$. 
As a result the Lam\'e soliton lattice energy has the simpler form
\begin{equation}
E_{SL}=2\,\sqrt{\displaystyle\frac{2\,p}{z_1\,-\,z_2}}\,
\left\{\left[z_1\,K(t)\,-\,(z_1\,-\,1)\,
\Pi\left(\displaystyle\frac{1\,-\,z_2}{z_1\,-\,z_2},\,t\right)\right]\,-\,
\displaystyle\frac{a^2\,K(t)}{2}\right\}\,.
\end{equation}

{\bf Case II.1 ($p<0$)} :
This produces {\em  type II} Lam\'e soliton lattice solution.
In this case the solution is in fact the same as that given by 
Eq. (\ref{latticeII.1}), except that now $z_1,z_2$ take the simpler 
values  $z_1={1}/{k^2}$, $z_2=1-a^2$. As a result the Lam\'e 
soliton lattice energy has the simpler form
\begin{equation}
E_{SL}=2\,\sqrt{\displaystyle\frac{2\,|p|}{z_1\,-\,z_2}}\,\left\{
\left[-(z_1\,-\,1)\,K(r)\,+\,z_1\,\Pi\left(-\,
\displaystyle\frac{z_2}{z_1\,-\,z_2},\,r\right)\right]\,-\,  
\,\displaystyle\frac{a^2\,K(r)}{2}\right\}\,.
\end{equation}

\subsection{Lam\'e Single Soliton Solutions}

We can now easily obtain the corresponding Lam\'e soliton solutions, either
by taking the appropriate limits of the Lam\'e soliton lattice solutions, or of the 
AL soliton solutions. 

{\bf Type I  ($p > 0$)} :
For example, considering $\Gamma =0$ in the expressions for the AL soliton 
solution as given by Eq. (\ref{singleI.1}), the corresponding Lam\'e 
one soliton solution turns out to be
\begin{equation}
\mbox{sn}(\phi,\,k) =\frac{1}{\sqrt{1+k'^2\,\mbox{sinh}^2(y^*)}}\, ,
\end{equation}
where $y^{*}=\sqrt{2\,p\,k^2}\,x$. The corresponding energy turns out to be
\begin{equation}
E=\sqrt{2\,p}\;\mbox{ln} \left(\frac{1+k}{1-k}\right)\, ,
\end{equation}
where we have used Eq. (\ref{energysingleI.1}).

{\bf Type II  ($p < 0$)} :
Considering  $\Gamma =0$ in the expressions for the AL soliton solution
as given by Eq. (\ref{singleII.1}) the corresponding Lam\'e soliton 
solution turns out to be
\begin{equation}
\mbox{sn}(\phi,\,k) =\frac{\mbox{sinh}(y^{*})}{\sqrt{k'^2+\mbox{sinh}^2(y^*)}}\, ,
\end{equation}
where $y^{*}=\sqrt{2\,|p|\,k^2\,k'^2}\,x$. The corresponding energy turns out to be
\begin{equation}
E=2\sqrt{2\,|p|}\;\mbox{arctan} \left(\frac{k}{k'}\right)\, ,
\end{equation}
where we have used Eq. (\ref{energysingleII.1}).

\section{Conclusion} 
We have presented the exact soliton lattice and single soliton solutions,
as well as their corresponding energies, for the Associated Lam\'e equation 
\cite{magnus,khare,ganguly} in various parameter regimes.  In appropriate 
limits we also obtained the single soliton and soliton lattice solutions 
of the Lam\'e equation. The topological and nontopological nature of the 
different solutions was discussed. As an illustration of Manton's method 
\cite{manton} we also computed, in a particular case, the asymptotic 
interaction energy  between these solitons.  In addition to their relevance 
in the study of nonlinear phenomena, these solutions provide valuable 
information about domain walls in field theory, materials and many physical 
systems. It would be worthwhile to analytically study the exact 
thermodynamical properties of these systems. 
\section{Acknowledgments} 

I.B. acknowledges support from the Swiss National Science Foundation.
This work was supported in part by the U.S. Department of Energy.

\end{document}